\documentclass[12pt]{article}

\usepackage{amsfonts}
\usepackage{amsmath}
\usepackage{epsfig}
\usepackage{latexsym}
\usepackage{graphicx}
\numberwithin{equation}{section}
\allowdisplaybreaks

\def\be{\begin{equation}}
\def\ee{\end{equation}}
\def\bea{\begin{eqnarray}}
\def\eea{\end{eqnarray}}

\def\half{{1\over 2}}

\def\del{\partial}
\def\nn{\nonumber}

\renewcommand{\thefootnote}{\fnsymbol{footnote}}

\begin{document}

\hfuzz=100pt
\title{{\Large \bf{Brown-Teitelboim Instantons and the First Law of Thermodynamics of (Anti) de Sitter Space}}}
\date{}
\author{Shinji Hirano$^{a,b}$\footnote{shinji.hirano@wits.ac.za}
  \\ \\
$^a${\small{\it School of Physics and Mandelstam Institute for Theoretical Physics}}
\\{\small{\it \& DST-NRF Centre of Excellence in Mathematical and Statistical Sciences (CoE-MaSS) }}
\\{\small{\it University of the Witwatersrand, WITS 2050, Johannesburg, South Africa}}\\ 
$^b${\small{\it Center for Gravitational Physics,
Yukawa Institute for Theoretical Physics}}
\\{\small{\it Kyoto University, Kyoto 606-8502, Japan }}\\
\\
}
\date{}

\maketitle
\centerline{}

\begin{abstract}

We study the instantons (or bounces) in the Brown-Teitelboim (BT) mechanism of relaxation of cosmological constant which is a cosmological version of the Schwinger mechanism. The BT mechanism is a false vacuum decay of (A)dS$_{d+1}$ (and $\mathbb{R}^{1, d}$) spaces via spontaneous nucleations of  spherical $(d-1)$-branes and thus ostensibly has bearings  on (A)dS$_{d+1}$/CFT$_d$ holography. 
In this paper we focus on the four-dimensional case, although the higher or lower-dimensional generalization is straightforward.
As is the case with pair productions near black hole and de Sitter horizons, we show that the BT instanton action for a membrane nucleation encodes the first law of thermodynamics of (Anti) de Sitter space. In particular, the membrane instanton precisely accounts for the change of entropy of (A)dS space before and after nucleation, in good accordance with AdS$_{d+1}$/CFT$_d$ in which the $(d-1)$-branes make up all degrees of freedom of AdS$_{d+1}$ space. 
In light of this lesser-known perspective presented here we also make remarks on (1) (A)dS/CFT and (2) complexity. For the complexity we observe that the Lorentzian bounce action may have close connection to complexity. 

\end{abstract}

\renewcommand{\thefootnote}{\arabic{footnote}}
\setcounter{footnote}{0}

\newpage

\section{Introduction}

\begin{figure}[h]
\centering
\hspace{-.7cm}
\includegraphics[width=2.in]{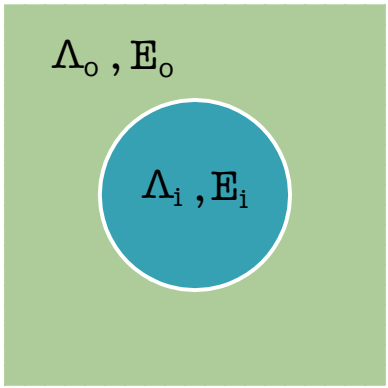}
\caption{The BT mechanism: With sufficient supply of energy from the flux, spherical membranes can be nucleated. The spacetime is divided into two by a spherical membrane indicated by the white circle. By nucleation of a bubble of a spherical membrane of charge $q$, the cosmological constant $\Lambda_{\rm o}$ and the 4-form flux $E_{\rm o}$ are reduced to $\Lambda_{\rm i}=\Lambda_{\rm o}-\kappa q\langle E\rangle<\Lambda_{\rm o}$ and $E_{\rm i}=E_{\rm o}-q<E_{\rm o}$ in the spacetime inside of the membrane, where $\langle E\rangle = E_{\rm o}-q/2$ is the average of the 4-form fluxes on the membrane. The bubble nucleation continues until the flux is reduced to a critical value $E_{c}$ at which point  there is not enough flux energy left to nucleate membranes and the true vacuum spacetime is an Anti de Sitter space.}
\label{fig:bubble0}
\end{figure}

The Brown-Teitelboim (BT) mechanism is a cosmological version of the Schwinger mechanism \cite{Schwinger:1951nm} in which the cosmological constant relaxes to a smaller value via nucleations of spherical membranes \cite{Brown:1987dd}.
This is a false vacuum decay of (A)dS$_{d+1}$ and $\mathbb{R}^{1, d}$ spaces \cite{Coleman:1977py, Coleman:1980aw}, more generally, via spontaneous nucleations of  spherical $(d-1)$-branes. (See Figure \ref{fig:bubble0} for an illustration of the BT mechanism.)
It is a beautiful dynamical mechanism and was proposed as a way to solve the cosmological constant problem.
However, in its original form, in order for the true vacuum spacetime to land in cosmological constants within the observational bound, the membrane charge needs to be extremely small compared to natural microphysics scales. Moreover, the decay rate is typically so small that the prolonged de Sitter expansion would leave the universe dead empty. These issues were revisited in \cite{Bousso:2000xa, Feng:2000if} and promising resolutions were proposed by embedding the BT mechanism into flux compactifications of String/M-theory.

In this paper we have nothing to add to the cosmological constant problem, but we instead discuss some other aspects of the BT mechanism.
Since the key ingredients are $(d-1)$-branes and the $(d+1)$-form flux, the BT mechanism ostensibly has bearings  on (A)dS$_{d+1}$/CFT$_d$ holography \cite{tHooft:1993dmi, Susskind:1994vu, Maldacena:1997re}. As an illustration of our points, we focus on the four-dimensional case, although the higher or lower-dimensional generalization is straightforward.\footnote{The two-dimensional case was studied, for example, in \cite{Frob:2014zka} albeit with very different objectives. }
In particular, we examine instanton (or bounce) solutions of the membrane nucleation first in the probe approximation and second, as done in BT's original works, by treating them as more fully-fledged gravitating domain walls. 
 We then show that, as is the case with pair productions near black hole and de Sitter horizons \cite{Parikh:1999mf, Parikh:1998mg, KeskiVakkuri:1996xp, Parikh:2002qh}, the BT instanton action encodes the first law of thermodynamics of (Anti) de Sitter space. In particular, the membrane instanton precisely accounts for the change of entropy of (A)dS space before and after nucleation, in good accordance with AdS$_{d+1}$/CFT$_d$ in which the $(d-1)$-branes make up all degrees of freedom of AdS$_{d+1}$ space. 
In light of the perspective the BT mechanism may offer we make further remarks on AdS/CFT as well as dS/CFT \cite{Strominger:2001pn, Witten:2001kn}. We also discuss complexity and make an observation that the Lorentzian bounce action may have close connection to complexity.

The organization of the paper is as follows: In Section \ref{BTinstantons} we provide a refreshing review on BT instanton (or bounce) solutions of the membrane nucleation (1) in the conformally flat metric of four-dimensional de Sitter space and (2) the dS$_3$ slice of four-dimensional de Sitter, flat and Anti de Sitter spaces. The former manifests itself as being the most intuitive as false vacuum decay and can be regarded as a direct and apparent higher dimensional generalization of the two-dimensional Schwinger mechanism in a uniform electric flux. The latter is simpler and more suitable and instrumental to discuss the BT mechanism. 
In Sections \ref{CFC} and \ref{dS3} we analyze membrane instantons in the probe approximation and in Section \ref{dS3gravity}, as done in original BT's works, we treat them as more fully-fledged gravitating domain walls.
As may be anticipated, the two analyses lead to essentially the same qualitative picture. 
In Section \ref{entropy} we discuss the thermodynamic interpretation of the BT instanton action and show that it obeys the first law of thermodynamics of (Anti) de Sitter space, as is the case with pair productions near black hole and de Sitter horizons.
In Section \ref{Discussions} we summarize our results and end the section with remarks on (A)dS/CFT and discussions on complexity.

\section{A refreshing review of BT instantons}
\label{BTinstantons}

We review the BT instantons of membrane nucleations in detail. In the original paper \cite{Brown:1987dd} as well as the literature that followed \cite{Feng:2000if, Donoghue:2003vs}, the BT instantons were analyzed in the static patch of de Sitter space. Here we provide a more intuitive exposition of the BT instantons by studying them in different coordinate systems of de Sitter space. In particular, we wish to convey that the probe approximation is very useful and sufficient to infer the essential results in the original works of Brown and Teitelboim in which they treated the membranes as gravitating domain walls beyond the probe approximation. 

The setup of the BT mechanism is as follows: The action consists of the Einstein gravity, the \lq\lq Maxwell theory'' of a (non-dynamical) three-form $C_3$ and membranes sourcing the four-form flux $F_4=dC_3$: 
\begin{align}
S=&{1\over 2\kappa}\int d^4x\sqrt{-g}\left(R-2\lambda_{\rm bare}-{2\kappa\over 2\cdot 4!}|F_4|^2\right)\nn\\
&\hspace{1.5cm}+{1\over\kappa}\int d^3x \sqrt{h}K+{1\over 3!}\int d^4x\del_{\mu_1}\left[\sqrt{-g}F^{\mu_1\cdots\mu_4}C_{\mu_2\cdots\mu_4}\right]
+S_{M2}
\label{Action}
\end{align}
where the signature convention is $(-,+,+,+)$, Newton's constant $\kappa=8\pi G$ and the kinetic term for the three-form field,  $|F_4|^2=F_{\mu_1\mu_2\mu_3\mu_4}F^{\mu_1\mu_2\mu_3\mu_4}$. The membrane action $S_{M2}$ is given by
\begin{align}
S_{M2}=-T_2\int_{M_3} d^3\sigma\sqrt{-\det g_{\rm ind}}+q\int_{M_3} C_3
\label{M2action}
\end{align}
where $T_2$ and $q$ are the membrane tension and charge, respectively, and $g_{\rm ind}$ is the induced metric on the membrane. Note that for this dynamical system to be well-defined, it is important to add the Gibbons-Hawking-York boundary term \cite{Gibbons:1976ue, York:1972sj} as well as that for the three-form \cite{Brown:1987dd, Duncan:1989ug} in the second line of \eqref{Action}.

The cosmological constant $\Lambda$ of the universe has two components, the bare cosmological constant $\lambda_{\rm bare}$ and the contribution from a uniform 4-form flux $E$. To be more precise, it has the form
\be
2\Lambda =2\lambda_{\rm bare}-{2\kappa\over 2\cdot 4!} |F_4|^2=2\lambda_{\rm bare}+\kappa E^2
\ee
where the 4-form is proportional the four-dimensional volume form
\be
F_4=dC_3= E dV_4\qquad\qquad{\rm and}\qquad\qquad |F_4|^2=-4! E^2\ ,
\ee
where $dV_4$ is the volume form of the de Sitter space. 
In string/M-theory compactifications the bare cosmological constant $\lambda_{\rm bare}$ is typically negative, because typical six/seven-dimensional internal manifolds have positive curvatures. 

\subsection{Instantons in conformally flat metric}
\label{CFC}

We first wish to study membrane nucleations in FRW universe. As we will see, among other choices of the charts, this manifests itself as the most intuitive picture of false vacuum decay in Coleman's original sense \cite{Coleman:1977py} and can be considered as the most direct and apparent higher-dimensional generalization of the two-dimensional Schwinger mechanism in a uniform electric flux \cite{Schwinger:1951nm}.

\begin{figure}[h]
\centering
\hspace{-.7cm}
\includegraphics[width=2.4in]{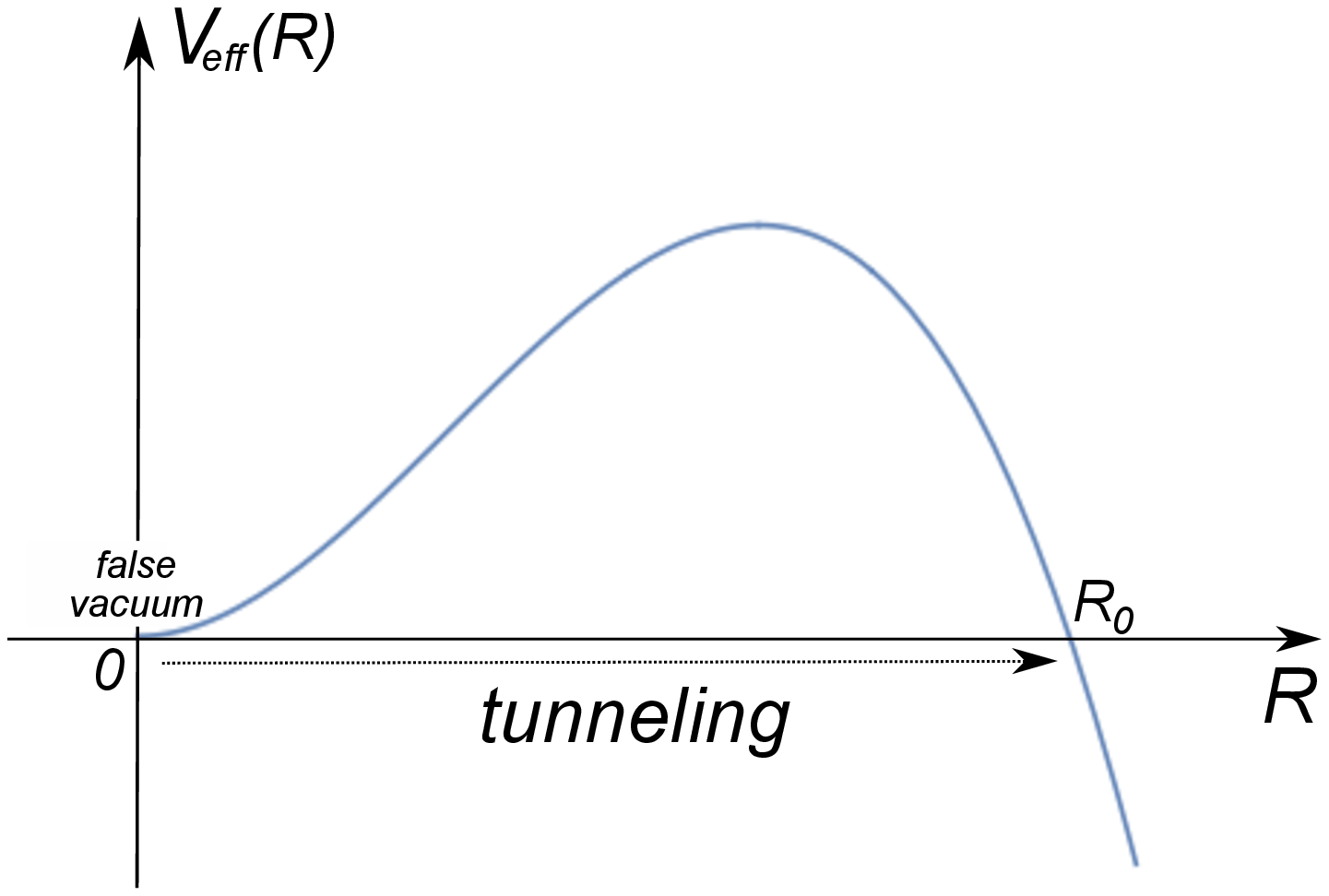}
\hspace{1.6cm}
\includegraphics[width=2in]{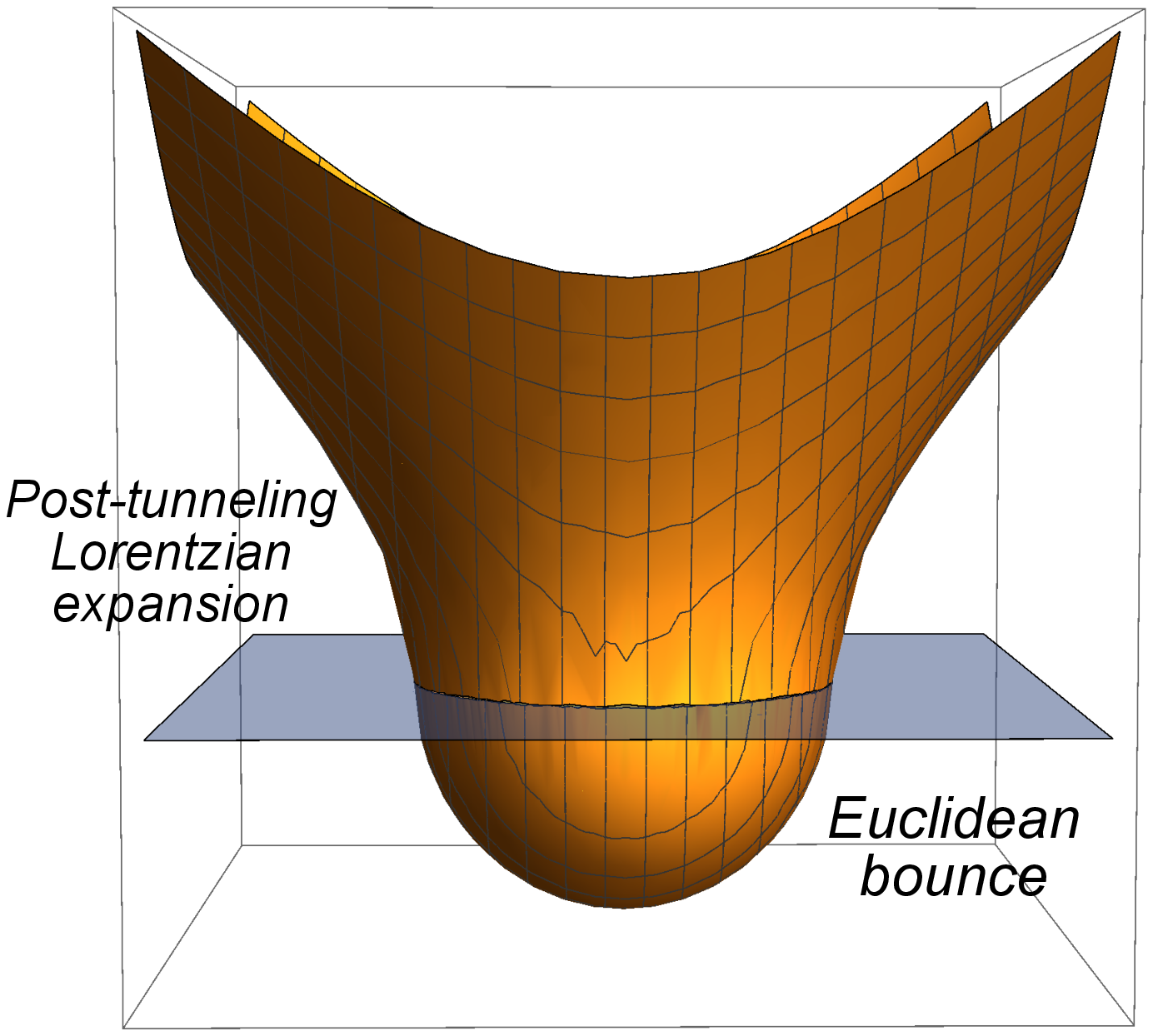}
\caption{The effective potential $V_{\rm eff}(R)$ on the left and the BT instanton (or bounce) on the right:
Tunneling through the potential barrier from the false vacuum at $R=0$, a spherical membrane of radius $R_0={3T_2H|\tau_0|\over qE}$ is nucleated at some time $\tau_0$ and expands out to the speed of light. The worldvolume geometry of the membrane instanton is $S^3$ in the Euclidean nucleation process and dS$_3$ in the post-tunneling Lorentzian expansion. }
\label{fig:Pot1}
\end{figure}

To be more precise, we look for membrane instantons in the spatially flat de Sitter universe
\be
ds^2=-dt^2 + e^{2H t}\left(dr^2 + r^2 d\Omega_2^2\right)\ ,
\ee
where the Hubble constant $H = \sqrt{\Lambda/3}$. It will prove most convenient to work with the conformal time $\tau=-H^{-1}e^{-Ht}<0$ in terms of which the metric becomes manifestly conformally flat
\be
ds^2={-d\tau^2+dr^2 + r^2 d\Omega_2^2\over H^2\tau^2}\ .
\label{CF}
\ee
In this coordinate system, the spherical membrane action \eqref{M2action} takes the form
\begin{align}
S_{M2}=\int {d\tau\over(H\tau)^3} d\Omega_2\left[-T_2R^2\sqrt{1-\dot{R}^2}-{qE\over 3H\tau} R^3\right]\ ,
\label{M2actionCF}
\end{align}
where we used the induced metric on a spherical membrane of radius $R(\tau)$
\be
ds_{M2}^2={1\over H^2\tau^2}\left[-\left(1-\dot{R}^2\right)d\tau^2+R^2d\Omega_2^2\right]
\ee
and the 3-form potential to which the membrane couples is given by
\be
C_3=-{1\over 3}E(H\tau)^{-4}  R^3d\tau\wedge d\Omega_2\ .
\ee
Note that the effective potential has the form
\be
V_{\rm eff}(R)={T_2\over |H\tau|^3}\left[R^2-{qE\over 3T_2H|\tau|}R^3\right]\ ,
\ee
where $\tau=-|\tau|$. The potential is plotted in the left panel of Figure \ref{fig:Pot1} and it provides a very intuitive picture of the membrane nucleation as Coleman's false vacuum decay. Moreover, the instanton solution is an apparent higher-dimensional generalization of that in the Schwinger mechanism in two dimensions. In fact, it is easy to check that the dS$_3$ worldvolume
\be
{\rm dS_3\,\,\,\, Lorentzian:}\qquad\quad
R(\tau)^2-(\tau-\tau_0)^2=R_0^2\qquad\quad\mbox{with}\qquad\quad R_0={3T_2H|\tau_0|\over qE}
\label{instantonCF}
\ee
solves the equation of motion of the membrane action \eqref{M2actionCF}
\be
{d\over d\tau}\left({R^2\dot{R}\over\tau^3\sqrt{1-\dot{R}^2}}\right)
+{1\over\tau^3}\left[2R\sqrt{1-\dot{R}^2}+{qE\over T_2H\tau} R^2\right]=0\ .
\ee
By the Wick-rotation $\tau= \tau_0 + i\tau_E$, we obtain the Euclidean membrane nucleation process
\be
{\rm S^3\,\,\,\, Euclidean:}\qquad\quad
R(\tau)^2+\tau_E^2=R_0^2
\label{EinstantonCF}
\ee
which is a $S^3$ and smoothly connected to the post-tunneling Lorentzian evolution \eqref{instantonCF} at $\tau_E=0$, or equivalently $\tau=\tau_0$, as depicted in the right panel of  Figure \ref{fig:Pot1}. This is the BT instanton (or bounce) in the conformally flat metric.

Note that there are four zero modes and one negative mode in a BT instanton. The zero modes appear since the $SO(4,1)$ isometry of the de Sitter space is broken to $SO(3,1)$ by the nucleation of the BT instanton. The instant of the membrane nucleation $\tau_0$ is one of the four zero modes associated with the broken scale invariance of the de Sitter space \eqref{CF}. The remaining three are the position $\vec{x}_0$ of the membrane in $\mathbb{R}^3$ associated with the broken translation invariance. The negative mode is the one which takes the radius away from $R_0$.  

We remark that the flat space limit can be reached by choosing $\tau_0=-H^{-1}$ and sending $H\to 0$ while keeping $t=-H^{-1}\ln(-H\tau)$ fixed. The (post-tunneling) instanton solution remains essentially the same
\be
{\rm dS_3\,\,\,\, Lorentzian:}\qquad\quad
R(t)^2-t^2=R_0^2\qquad\quad\mbox{with}\qquad\quad R_0={3T_2\over qE}\ .
\label{instantonFlat}
\ee
Note that this is perfectly analogous to the radius of an electron-positron pair nucleation, $r_0={m\over eE}$, in the Schwinger mechanism. In this case the Wick-rotation to the Euclidean nucleation process is simply $t=it_{\rm E}$.

\subsubsection{The nucleation rate}
\label{CFdecayrate}

We now wish to calculate the nucleation rate of the spherical membrane. For that purpose, it is convenient to introduce the rescaled variable $x=\tau/\tau_0$ and then Wick-rotate as $x=iy+1$. 
As pioneered in Coleman's work \cite{Coleman:1977py}, the nucleation or decay rate $\Gamma$ associated with a bounce solution is given by the imaginary part of the energy ${\rm Im}E=\Gamma/2\propto e^{-S_0}$, where $S_0$ is the Euclidean bounce (or instanton) action and the energy acquires the imaginary part due to the negative mode of the bounce. 

The Euclidean instanton action can be found as
\begin{align}
S_{M2}({\rm INST})&=-i{qEV_{S^2}\over 3H^4}\int_{-R_0/|\tau_0|}^{+R_0/|\tau_0|}  dy\frac{y \left(i  y-(R_0/\tau_0)^2 \right) \sqrt{(R_0/\tau_0)^2-y^2}}{ (iy+1)^4 }\nn\\
&={2\pi^2 qE\over 3H^4} {\left(\sqrt{1+c^2}-1\right)^2\over \sqrt{1+c^2}}\ ,\label{InstactionCF}
\end{align}
where we used the volume of unit two-sphere, $V_{S^2}=4\pi$, and defined $c=R_0/|\tau_0|=3T_2H/(qE)$.
Note that in the flat space limit $H\to 0$, the instanton action $S_{M2}({\rm INST})\to {27\pi^2T_2^4\over 2(qE)^3}$. This is perfectly a sensible result and to be compared with the the case of the Schwinger mechanism, $S({\rm INST})={\pi m^2\over eE}$.
Another limit of interest may be $E\to 0$. In this limit the instanton action $S_{M2}({\rm INST})\to {2\pi^2T_2\over H^3}$ and this can be interpreted as the rate of thermal membrane productions in the de Sitter space \cite{Gibbons:1976ue}.

As will be elaborated in the next section, although the conformally flat metric provides the most intuitive picture of membrane instantons as false vacuum decay, it is not the most convenient description for the full realization of the BT mechanism. 
We thus seek an alternative description of membrane instantons in a different coordinate system.

\subsection{Instantons in dS$_3$ slice}
\label{dS3}

The dS$_3$ slices of four-dimensional de Sitter, flat and Anti de Sitter spaces are most suitable and instrumental to study the BT mechanism. Namely, the spacetime in the BT mechanism is divided into two across the membrane instanton, i.e. a domain wall. The spacetime outside has a greater cosmological constant than that of the spacetime inside. (See Figure \ref{fig:bubble0}): At early stages of the vacuum decay, both spacetimes are de Sitter. As the process progresses, the spacetime inside becomes a flat spacetime and further decays into an AdS space. At late stages it is possible that the spacetime outside becomes flat or AdS, while the spacetime inside is AdS with a smaller (more negative) cosmological constant.
The dS$_3$ slices are instrumental in the sense that patching the two spacetimes across the domain wall becomes very straightforward.
On the other hand,  in the conformally flat metric, it is not obvious how one should go from the dS or flat space to the AdS space.

We thus look for BT instantons in the dS$_3$ slices on which the Lorentzian instantons live: 
\begin{align}
ds_{\rm dS}^2&=d\chi^2 +H^{-2}\sin^2(H\chi) ds_{dS_3}^2\label{dS3slicedS}\ ,\\
ds_{\rm flat}^2&=dr^2 + r^2ds_{dS_3}^2\ ,\\
ds_{\rm AdS}^2&=d\rho^2 +R_{AdS}^2\sinh^2(\rho/R_{AdS}) ds_{dS_3}^2\ .
\label{dS3slice}
\end{align}
We find it most convenient to work with the dS$_3$ metric in the coordinates
\be
ds_{dS_3}^2=-d\eta^2+\cosh^2\eta d\Omega_2^2\ .
\ee
Note that the dS$_3$ slices smoothly connect dS$_4$ to 4d flat space to AdS$_4$ by taking the limit $H\to 0$ and then analytically continuing $H^{-1}=iR_{AdS}$.

\begin{figure}[h]
\centering
\includegraphics[width=2.4in]{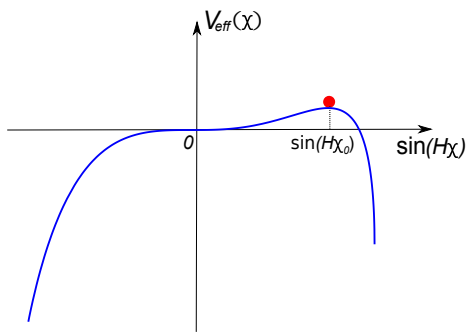}
\caption{The effective potential $V_{\rm eff}(\chi)$ for the spherical membrane in the dS$_3$ slice of dS$_4$:
In contrast to the case of the conformally flat metric, the BT instanton sits at the hilltop of the potential.}
\label{fig:Pot2}
\end{figure}

For the dS$_3$ worldvolume which is of our interest, the membrane action yields
\begin{align}
\hspace{-.5cm}
S_{M2} = -\int {dV_{dS_3}\over H^3}\!\left[T_2H\sin^2(H\chi)\sqrt{{\sin^2(H\chi)\over H^2}-\dot{\chi}^2}-{qE\over 3H}\left(\cos^3(H\chi)-3\cos(H\chi)\!+2\right)\right]\label{M2dS3}
\end{align}
where $\dot{\chi}=d\chi/d\eta$ and we fixed the gauge of the 3-form potential such that the membrane action vanishes when the size of the membrane is zero. 
The effective potential has the form
\be
V_{\rm eff}(\chi)={1\over H^3}\left[T_2\sin^3(H\chi)-{qE\over 3H}\left(\cos^3(H\chi)-3\cos(H\chi)\!+2\right)\right]\ .
\ee
The shape of the potential is plotted in Figure \ref{fig:Pot2}. As we will show, 
the BT instanton corresponds to a ``particle'' sitting at the hilltop of the potential. The maximum of the potential can be found from 
\begin{align}
V'_{\rm eff}(\chi)={1\over H^3}\sin^2(H\chi)\left(3T_2H\cos(H\chi)-qE\sin(H\chi)\right)=0\ .
\end{align}
The radius of the BT instanton thus reads
\begin{align}
\tan(H\chi_0)={3T_2H\over qE}\
\qquad\Longrightarrow\qquad
R_0\equiv H^{-1}\sin(H\chi_0)=\frac{{3T_2\over qE}}{\sqrt{1+\left({3HT_2\over qE}\right)^2}}\ .\label{radiusdS3}
\end{align}
Note that the maximal size at which a membrane can be nucleated is the Hubble radius $R_H=H^{-1}$.

After Wick-rotating the time $\eta=-i\eta_{\rm E}$, the dS$_3$ becomes a unit $S^3$ and we find the value of the Euclidean action to be
\begin{align}
S_{M2}(\chi_0)={2\pi^2 qE\over 3 H^4}\frac{(\sqrt{1+c^2}-1)^2}{\sqrt{1+c^2}}
\label{InstactiondS3}
\end{align}
where $c={3T_2H\over qE}$. This is exactly the same as the BT instanton action \eqref{InstactionCF} in the conformally flat metric, as anticipated from general covariance.
Note that being on the hilltop, there is trivially a negative mode as required for a decay channel.

It is now straightforward to generalize the BT instanton in the de Sitter space to those in the flat and Anti de Sitter spaces. We simply take the $H\to 0$ limit for the flat space and analytically continue $H^{-1}=iR_{AdS}$ for the Anti de Sitter space. From \eqref{radiusdS3} we find 
\begin{align}
R_0({\rm flat})={3T_2\over qE}\ ,\qquad\qquad\quad
R_0({\rm AdS})=\frac{{3T_2\over qE}}{\sqrt{1-\left({3T_2\over qER_{AdS}}\right)^2}}\ .
\end{align} 
In the AdS case, in particular, observe that there is a critical value of the 4-form flux $E_c$ at which the nucleation radius becomes infinity:
\begin{align}
E> E_c={3T_2\over qR_{AdS}}
\label{probecritical}
\end{align}
Since the constant $c$ in the Euclidean action \eqref{InstactiondS3} with $H^{-1}=iR_{AdS}$ becomes the imaginary $i$ at the critical 4-form flux $E_c$, the Euclidean action diverges and the decay rate vanishes. As discussed in \cite{Brown:1987dd}, this means that the false vacuum decay stops when the 4-form flux reduces to the critical value.

We remark that the critical 4-form flux corresponds to BPS branes with the tension-to-charge ratio
\be
{T_2\over q_{\rm\scriptscriptstyle BPS}}={ER_{AdS}\over 3} 
\label{BPS}
\ee
as discussed in \cite{Maldacena:1998uz}. Note that this is when the maximum of the potential goes down and degenerates to zero. It is interesting to observe that although the nucleation rate is zero, if they were nucleated, the BPS branes must appear at infinity, that is, at the boundary of the AdS space. This is in accordance with holography in which the BPS branes play the role of the holographic screen at the boundary.

The situation analogous to this criticality in the dS case is when the nucleation radius \eqref{radiusdS3} becomes maximal, i.e. the Hubble radius $R_H$. This corresponds to the zero-charge limit $q\to 0$, that is, when branes are neutral.
We will come back to this point and discuss it a little further  in Section \ref{Discussions}.

\subsection{Gravitating membrane domain walls in dS$_3$ slice}
\label{dS3gravity}

In the previous sections we studied BT instantons in the probe approximation. To be complete, as analyzed in BT's original papers \cite{Brown:1987dd}, we now go beyond the probe approximation to treat them as more fully-fledged gravitating domain walls and include the backreaction. As we will see, although there are some corrections to the probe results, the qualitative picture remains essentially the same.

On and across the domain wall there are three conditions to be satisfied, which follow from the equations of motion for the action \eqref{Action}. One is the Israel junction conditions across the domain wall \cite{Israel:1967zz},
\begin{align}
\hspace{-3.5cm}
{\rm (Junction\,\, conditions):}\quad\qquad
K_{ab}({\rm out})-K_{ab}({\rm in})={\kappa T_2\over 2}(g_{\rm ind})_{ab}
\label{Israel}
\end{align}
where the extrinsic curvature $K_{ab}=-\half(\nabla_a n_b+\nabla_b n_a)$ with a unit vector $n_a$ normal to the domain wall surface.
The second is the change of the  4-form flux across the domain wall
\be
\hspace{-4.5cm}
{\rm (Flux\,\, condition):}\quad\qquad\qquad
E_{\rm i}=E_{\rm o}-q
\label{fluxchange}
\ee
where the inside flux $E_{\rm i}$ is reduced by $q$ relative to the outside flux $E_{\rm o}$.
The last is the equations of motion for the membrane embedding which yield
\be
\hspace{-4.5cm}
{\rm (Membrane\,\, EOM):}\quad\qquad\qquad
\langle K\rangle =-{q\over T_2}\langle E\rangle \label{M2EOM1}
\ee
where $\langle K\rangle$ and $\langle E\rangle$ are the extrinsic scalar curvature and the 4-form flux on the domain wall, respectively, and taken to be the averages
\be
\langle K\rangle=\half\left(K({\rm out})+K({\rm in})\right)\ ,\qquad
\langle E\rangle=\half\left(E_{\rm o}+E_{\rm i}\right)\ .
\ee
In terms of the dS$_3$ coordinates \eqref{dS3slice} for the de Sitter space, the equations \eqref{Israel} and \eqref{M2EOM1} read
\begin{align}
&-\half H_{\rm o}^{-1}\sin(2H_{\rm o}\chi_{\rm o})+\half H_{\rm i}^{-1}\sin(2H_{\rm i}\chi_{\rm i})={\kappa T_2\over 2}
\langle H^{-2}\sin^2(H\chi_0)\rangle\ ,\label{junction2}\\
&-\half H_{\rm o}^{-1}\sin(2H_{\rm o}\chi_{\rm o})-\half H_{\rm i}^{-1}\sin(2H_{\rm i}\chi_{\rm i})=-{2q\over 3T_2}\langle E\rangle R^2\ ,
\label{M2EOM2}
\end{align}
where the subscripts o and i indicate outside and inside of the domain wall and $R$ is the membrane radius given by
\be
R=\langle H^{-1}\sin(H\chi_0)\rangle=H_{\rm o}^{-1}\sin(H_{\rm o}\chi_{\rm o})
=H_{\rm i}^{-1}\sin(H_{\rm i}\chi_{\rm i})\ .\label{radiusconsistency}
\ee
These two equations are solved by\footnote{One can check that these are consistent with the relations among the radii \eqref{radiusconsistency} by noticing that
\be
\cos^2(H_{\rm o}\chi_{\rm o})-\cos^2(H_{\rm i}\chi_{\rm i})=R^2(H_{\rm i}^2-H_{\rm o}^2)=-{\kappa q\langle E\rangle\over 3}R^2\ .\nn
\ee}
\begin{align}
\cos(H_{\rm o}\chi_{\rm o})&={R\over 2}\left({2q\langle E\rangle\over 3T_2}-{\kappa T_2\over 2}\right)\ ,\label{sol1}\\
\cos(H_{\rm i}\chi_{\rm i})&={R\over 2}\left({2q\langle E\rangle\over 3T_2}+{\kappa T_2\over 2}\right)\ .\label{sol2}
\end{align}
From \eqref{radiusconsistency} -- \eqref{sol2} we find the radius
\begin{align}
R=\frac{{3T_2\over q\langle E\rangle}}{\sqrt{\left(1-{3\kappa T_2^2\over 4q\langle E\rangle}\right)^2+\left({3H_{\rm o}T_2\over q\langle E\rangle}\right)^2}}=\frac{{3T_2\over q\langle E\rangle}}{\sqrt{\left(1+{3\kappa T_2^2\over 4q\langle E\rangle}\right)^2+\left({3H_{\rm i}T_2\over q\langle E\rangle}\right)^2}}\ .\label{radiusBT}
\end{align}
Note that this radius is exactly the same as that found by BT in the static patch of de Sitter space \cite{Brown:1987dd}.
Comparing this to the probe result $R_0$ in \eqref{radiusdS3}, we observe that the $1$ in the denominator of $R_0$ is shifted by $\mp{3\kappa T_2^2\over 4q\langle E\rangle}$:
\be
1\to 1\mp {3\kappa T_2^2\over 4q\langle E\rangle}\ .
\ee
This is the gravitational effect due to the backreaction of the membrane to the spacetime. Had there not been the Israel junction conditions \eqref{Israel} which account for the backreaction, there would not have been this shift.

As much in the same way as in the probe approximation, it is straightforward to generalize this result to the flat and AdS cases:
\begin{align}
R_0({\rm flat})=\frac{{3T_2\over q\langle E\rangle}}{\left|1-{3\kappa T_2^2\over 4q\langle E\rangle}\right|}\ ,\qquad\qquad\quad
R_0({\rm AdS})=\frac{{3T_2\over q\langle E\rangle}}{\sqrt{\left(1-{3\kappa T_2^2\over 4q\langle E\rangle}\right)^2-\left({3T_2\over q\langle E\rangle R_{AdS}}\right)^2}}\ ,
\label{radiusAdS}
\end{align}
where the AdS radius $R_{AdS}$ is the one outside of the membrane. 
Note that for the membrane nucleation to happen in the Anti de Sitter space, the 4-form flux must be in the range
\be
\langle E\rangle> {3\kappa T_2^2\over 4q}+ {3T_2\over qR_{AdS}}\qquad\qquad{\rm or}\qquad\qquad
\langle E\rangle< {3\kappa T_2^2\over 4q}- {3T_2\over qR_{AdS}}
\label{radiusrange}
\ee
provided that the flux is positive $\langle E\rangle=E_{\rm o}-{q\over 2} >0$. 
Since Newton's constant $\kappa$ is small, it is typically the case that ${3\kappa T_2^2\over 4q}\ll {3T_2\over qR_{AdS}}$. 
Thus the second inequality in \eqref{radiusrange} may be ignored and we have the critical value of the 4-form flux
\be
\langle E\rangle> \langle E_c\rangle=E_c+{3\kappa T_2^2\over 4q}\ ,
\ee
where we see the shift of the probe value $E_c$ in \eqref{probecritical} by the gravitational effect ${3\kappa T_2^2\over 4q}$. This is exactly the same as the critical flux discussed in \cite{Brown:1987dd}.

\subsection{The decay rate}
\label{dS3decayrate}

The decay rate of a false vacuum or the bubble nucleation rate $\Gamma$ is simply given by
\be
\Gamma\sim e^{-S_E({\rm INST})+S_E({\rm BKG})}
\label{decayrategravity}
\ee
where $S_E({\rm INST})$ and $S_E({\rm BKG})$ are the instanton and background actions, respectively, which are the Euclidean continuation of the action \eqref{Action} evaluated on the spacetimes after and before the membrane nucleation. Thus the nonvanishing contributions can come only from the domain wall and the spacetime inside. 

As in \cite{Brown:1987dd, Feng:2000if, Donoghue:2003vs}, the Euclidean continuation of the action \eqref{Action} at on-shell simplifies to
\begin{align}
S_E({\rm INST})=&-{1\over \kappa}\int_{{\rm in}/{\rm out}} d^4x\sqrt{-g}\Lambda+{1\over\kappa}\int_{{\rm in}/{\rm out}} d^3x \sqrt{h}K\ .
\end{align}
Both the bulk and boundary gravitational terms have contributions from the inside and outside of the spacetime.
To arrive at this form, we used the 3-form equation which converts the boundary and membrane terms into a bulk contribution and the trace of the Einstein equation which, in particular, cancels the membrane mass contribution.
Subtracting the background, this gives the Euclidean action
\begin{align}
\hspace{-.15cm}
\Delta S_{E}\equiv S_E({\rm INST})-S_E({\rm BKG})={1\over\kappa}\left[-V_{\rm EdS}(H_{\rm i})\Lambda_{\rm i}-{3V_{M2}(R)\over R}\cos(H_{\rm i}\chi_{\rm i})\right]-({\rm i}\to {\rm o})\ ,
\label{BTinstantonaction}
\end{align}
where the Euclidean de Sitter (i.e. $S^4$) volume is $V_{\rm EdS}(H)={2\pi^2\over 3H^4}\left(\cos^3(H\chi)-3\cos(H\chi)+2\right)$ and the $S^3$ membrane volume $V_{M2}(R)=2\pi^2R^3$. Using \eqref{radiusconsistency} and $\Lambda=3H^2$, we find that 
\begin{align}
\Delta S_{E}&={4\pi^2\over \kappa}\left[H_{\rm i}^{-2}\left(\cos(H_{\rm i}\chi_{\rm i})-1\right)
-H_{\rm o}^{-2}\left(\cos(H_{\rm o}\chi_{\rm o})-1\right)\right]\nn\\
&={4\pi^2\over \kappa}\left[H_{\rm i}^{-2}\left(\sqrt{1-(RH_{\rm i})^2}-1\right)
-H_{\rm o}^{-2}\left(\sqrt{1-(RH_{\rm o})^2}-1\right)\right]\ .
\label{instantonactionbackreaction}
\end{align}
This again is exactly the same result as that of BT in the static patch, as anticipated from general covariance.

\begin{figure}[h]
\centering
\hspace{-.7cm}
\includegraphics[width=4.0in]{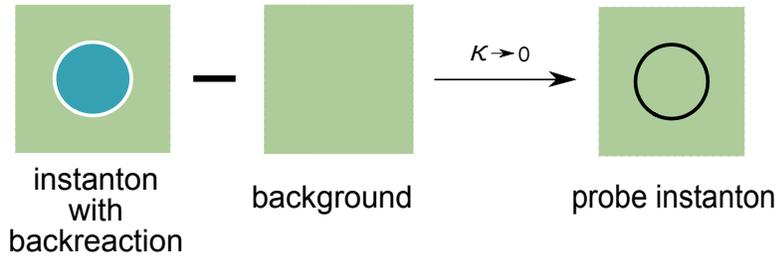}
\caption{An illustration of the decay rate computation: The decay rate of the fully-fledged gravitating membrane instanton is computed by evaluating the gravity action \eqref{Action} over the entire spacetime (with the background subtraction). In the probe limit $\kappa\to 0$, the decay rate is computed by evaluating the probe action \eqref{M2action} localized on the bubble indicated by the black circle.}
\label{fig:decayrate}
\end{figure}

To compare this action with the probe result, we expand it for a small $\kappa$. As expected, this reduces precisely to the probe instanton action
\begin{align}
\Delta S_{E}=S_{M2}(\chi_0)+O(\kappa)\ ,
\label{nogravityinstanton}
\end{align}
where we used \eqref{radiusBT} and $H_{\rm i}^2=H_{\rm o}^2-{\kappa q\langle E\rangle\over 3}$, and  $S_{M2}(\chi_0)$ is the probe instanton action \eqref{InstactiondS3} with the flux $E$ being replaced by the average $\langle E\rangle$ and the Hubble constant $H_{\rm o}$. An illustration of the computational schemes and results of the decay rate is given in Figure \ref{fig:decayrate}. 

As discussed by Brown and Teitelboim in their original papers \cite{Brown:1987dd}, the decay does not stop at zero cosmological constant but continues into an Anti de Sitter space. If the 4-form flux is at or below the critical value when the spacetime outside is de Sitter or flat and the inside spacetime is Anti de Sitter space, the nucleation process stops. If the 4-form flux is above the critical value when the spacetime inside is Anti de Sitter, it continues to decay to an Anti de Sitter space of a larger curvature until the flux reaches its critical value.
The emphasis of BT is that there is a large region in the parameter space which can achieve negative but very small cosmological constants in the true vacuum so that the BT mechanism can be viable for resolving the cosmological constant problem.

\section{The first law of thermodynamics of (Anti) de Sitter space}
\label{entropy}

Since membrane nucleations mediate the decay of de Sitter and Anti de Sitter spaces, the decay rate may encode the thermodynamic property of these spaces. More precisely, as is the case with pair productions near black hole and de Sitter horizons \cite{Parikh:1999mf, Parikh:1998mg, KeskiVakkuri:1996xp, Parikh:2002qh}, the decay rate in our approximation is expected to have the interpretation
\be
\Gamma\sim e^{-2S({\rm INST})}=e^{\Delta S-\beta \Delta{\cal E}}\ ,
\ee
where the instanton action $S({\rm INST})$ is either $S_{M2}({\rm INST})$ or $\Delta S_{E}$ with or without the backreaction. 
$\Delta S$ is the change of entropy of the dS or AdS space and $\Delta{\cal E}$ is the energy of a ``particle'' in the {\it thermal} ``pair production'' if it occurs. We will elaborate on these points below.

\subsection{de Sitter thermodynamics in the probe limit}
\label{dSentropyProbe}

We wish to understand the relation between the membrane nucleation and the thermodynamic property of (Anti) de Sitter space.
For this purpose we first analyze the instanton action in the probe limit when the changes of the cosmological constant and the flux are small.
Namely, we consider the case when the charge $q$ is small and thus the size of the membrane \eqref{radiusdS3} is close to the Hubble radius $R_H=H^{-1}$, the maximal possible size at which the membrane can be nucleated.

We thus expand the probe instanton action \eqref{InstactiondS3} for a small $q$, which indeed yields\footnote{To be more precise, a small $q$ means that $q\ll {3T_2H\over E}, {3H^2\over \kappa E}$.}
\begin{align}
S_{M2}(\chi_0)=\frac{2 \pi ^2T_2}{H^3}-\frac{4\pi^2 qE }{3 H^4}+O(q^2)
=\half\left(\beta \Delta{\cal E}-\Delta S_{dS}\right)+O(q^2)\ .
\label{ThermalInterpretationInstAction}
\end{align}
where $\beta=2\pi H^{-1}$ is the inverse temperature of the de Sitter space.
(1) $\Delta{\cal E}=2\pi T_2H^{-2}$ is half the energy of an $S^2$ membrane at the Hubble radius $R_H=H^{-1}$. Thus when the $S^2$ membrane is viewed as a dipole, it is the energy of a ``particle'' or an ``antiparticle'' of the thermally produced ``pair''.
(2) $\Delta S_{dS}$ is the change of the de Sitter entropy before and after the membrane nucleation.  
In more detail, it is most convenient to consider the static patch of the de Sitter space
\be
ds_{dS}^2=-(1-H^2r^2)dt^2 +{dr^2\over 1-H^2r^2}+r^2d\Omega_2^2\ .
\label{staticpatch}
\ee
The area of the cosmological horizon at $r=H^{-1}$ yields the entropy
\be
S_{dS}={A_{H}\over 4G}={\pi\over G H^2}\ .
\ee
Then the change of the entropy before and after the membrane nucleation can be found as
\be
\Delta S_{dS}={\pi\over G}\left({1\over H_{\rm i}^2}-{1\over H_{\rm o}^2}\right)
={8\pi^2 q\langle E\rangle\over 3H_{\rm o}^4}+O(\kappa, q^2)
\label{entropydiff}
\ee
where we used $H_{\rm i}^2=H_{\rm o}^2-{\kappa q\langle E\rangle\over 3}$ with $\kappa =8\pi G$. For a small $q$ we can replace $H_{\rm o}$ by $H$ and $\langle E\rangle$ by $E$. This proves the thermodynamic interpretation of \eqref{ThermalInterpretationInstAction}.

Since the probe approximation is the limit in which gravitational interactions are switched off, the probe instanton action \eqref{InstactiondS3} does not depend on Newton's constant $\kappa$ and cannot, in principle, account for the subleading corrections in the change of  entropy \eqref{entropydiff}. Thus the $O(q^2)$ corrections in \eqref{ThermalInterpretationInstAction} should be considered as part of the energy of the thermally nucleated ``particle''.

\subsection{de Sitter thermodynamics beyond the probe limit} 
\label{dSentropy}

The instanton action \eqref{instantonactionbackreaction} goes beyond the probe limit and takes into account the gravitational effects. First,
observe that the change of the dS entropy is fully encoded in this action:
\begin{align}
\Delta S_{E}=-\half\Delta S_{dS}+{4\pi^2\over \kappa}\left[H_{\rm i}^{-2}\sqrt{1-(RH_{\rm i})^2}
-H_{\rm o}^{-2}\sqrt{1-(RH_{\rm o})^2}\right]\ .
\end{align}
For the thermodynamic interpretation we wish to identify the second difference term with
\be
\beta_{\rm i} {\cal E}_{\rm i}-\beta_{\rm o} {\cal E}_{\rm o}={8\pi^2\over \kappa}\left[H_{\rm i}^{-2}\sqrt{1-(RH_{\rm i})^2}
-H_{\rm o}^{-2}\sqrt{1-(RH_{\rm o})^2}\right]\ ,
\label{energybackreacted}
\ee
where $\beta=2\pi H^{-1}$. 
In fact, it is straightforward to see that when expanded for a small $\kappa$ and then for a small $q$, this contribution reduces to
\begin{align}
{\rm RHS}\,\,{\rm of}\,\,\eqref{energybackreacted}=\beta_{\rm o}\left[\left(2\pi T_2H_{\rm o}^{-2}+O(q^2)\right)+\left({\pi\kappa T_2 q\langle E\rangle\over 2H_{\rm o}^4}+O(\kappa q^2)\right)+O(\kappa^2)\right]\ .
\end{align}
Thus, as anticipated from \eqref{nogravityinstanton} and \eqref{ThermalInterpretationInstAction},  the leading order contribution is $\beta\Delta{\cal E}$ of the probe limit \eqref{ThermalInterpretationInstAction}.

Note that in the static patch \eqref{staticpatch} the energy ${\cal E}$ an observer sees is related to the proper energy ${\cal E}_p$ by
\be
{\cal E} = \sqrt{1-(RH)^2}\,{\cal E}_p
\ee
at some fixed radius $r=R$. Then we identify the proper energy ${\cal E}_p=1/(2GH)$ in \eqref{energybackreacted} which is half the {\it thermal} energy of de Sitter space, ${\cal E}_p=\half{\cal E}_{th}$.\footnote{From $d{\cal E}_{th}=TdS_{dS}$ with $T=H/(2\pi)$, it can be inferred that ${\cal E}_{th}=1/(GH)$.} This reads
\be
{\cal E}_{\rm i, o}=\half({\cal E}_{th})_{\rm i,o}\sqrt{g_{00}(R)_{\rm i, o}}\ .
\ee
The factor of $\half$ again corresponds to the picture of a spherical membrane as a dipole and is associated with the energy of a ``particle'' or an ``antiparticle'' in the thermal ``pair production''. 
We thus conclude that
\begin{align}
\Delta S_{E}=\half\left[-\Delta S_{dS}+\left({\beta_{\rm i}\over 2}({\cal E}_{th})_{\rm i}\sqrt{g_{00}(R)_{\rm i}}-{\beta_{\rm o}\over 2}({\cal E}_{th})_{\rm o}\sqrt{g_{00}(R)_{\rm o}}\right)\right]\ .
\end{align}
As noted above, the difference of the thermal energy as seen by an observer at the leading order is the energy of a hemisphere membrane in the thermal ``pair production'' at the Hubble radius.

\subsection{The Anti de Sitter case}
\label{AdSentropy}

The AdS case is more subtle than the dS case in the following sense: (1) The entropy of the Anti de Sitter space we consider here is that of the hyperbolic (or topological) black hole which is isometric to the AdS space \cite{Emparan:1999gf}. The area is divergent and needs to be regularized. To extract the finite contribution, we use the regularized area of hyperbolic spaces prescribed in \cite{Maldacena:2012xp}.
(2) There is no membrane nucleation below the value of the BPS charge $q_{\rm\scriptscriptstyle BPS}={3T_2\over ER_{AdS}}$, and thus it may not make sense to consider  small $q$ expansions of the instanton action. Instead, what really is analogous to the dS case is to consider a small change from the maximal membrane nucleation which occurs at the charge $q_{\rm\scriptscriptstyle BPS}$ in the AdS case. Namely, we expand the instanton action \eqref{InstactiondS3} with $H^{-1}=iR_{AdS}$ and $q=q_{\rm\scriptscriptstyle BPS}+\delta q$ for a small $\delta q$.

The small $\delta q$ expansions yield
\begin{align}
S_{M2}(\chi_0)=&\frac{2\pi ^2T_2R_{AdS}^3}{\epsilon}-4\pi ^2T_2R_{AdS}^3\nn\\
&\hspace{1cm}+2 \pi ^2 \delta qER_{AdS}^4\left({1\over \epsilon}-{1\over 3\epsilon^3}\right)
-{4\over 3}\pi ^2 \delta qER_{AdS}^4
+O(\epsilon, \delta q^2)\ ,
\label{AdSinstanton}
\end{align}
where 
\be
\epsilon=\sqrt{1-\left(\frac{3T_2}{q_{\rm\scriptscriptstyle BPS} ER_{AdS}}\right)^2}\to 0\ .
\ee
The divergences as $\epsilon\to 0$ are the ``UV divergences'' coming from the boundary of AdS, since the membrane instanton is nucleated at the boundary when $q=q_{\rm\scriptscriptstyle BPS}$. Although the actual decay rate is zero due to these divergences, we shall now see that the finite part can still be interpreted as the first law of thermodynamics.

Meanwhile, the 4d hyperbolic (or topological) black hole takes the form 
\begin{align}
ds^2=-\left((r/R_{AdS})^2-1\right)dt^2+{dr^2\over (r/R_{AdS})^2-1}+r^2dH_2^2
\label{topBH}
\end{align}
which is isometric to the AdS$_4$ space. The entropy of the 4d hyperbolic black hole is thus given by 
\be
S_{AdS}={R_{AdS}^2V_{H_2}\over 4G}\ ,
\ee
Since the volume of the 2d hyperbolic space $H_2$ is infinite, so is the entropy. However,  
using the regularized volume for the unit 2d hyperbolic space $V_{H_2}=-\half V_{S^2}$ \cite{Maldacena:2012xp}, we can extract the finite change of the entropy which reads
\be
\Delta S_{AdS}=-{4\pi^2\over\kappa}\left(R_{AdS,{\rm i}}^2-R_{AdS,{\rm o}}^2\right)={4\over 3}\pi^2\delta q\langle E\rangle R_{AdS,{\rm o}}^2+O(\kappa, \delta q^2)\ ,
\ee
where we used $3R_{AdS, {\rm o,i}}^{-2}=-\lambda_{\rm bare}-\half\kappa E_{\rm o,i}^2$.
We thus find that the finite part of the decay rate can be expressed as
\be
\Gamma_{\rm fin}\sim e^{-2\left(S_{M2}(\chi_0)\right)_{\rm fin}}=e^{2\Delta S_{AdS}-\beta(2\Delta{\cal E})}\ ,
\label{1stlawAdS}
\ee
where $\beta=2\pi R_{AdS}$ and the energy of a hemisphere membrane $\Delta{\cal E}=-2\pi T_2R_{AdS}^2$. This is the finite part of the energy $2\pi T_2R_{AdS}^2(1/\epsilon -1)$ of a hemisphere membrane at the boundary, as can be inferred from \eqref{AdSinstanton}.

Note that the entropy and the energy might look twice as much as expected. However, we recall that there are two topological black holes in the global AdS. Since our result of the decay rate in the dS$_3$ slice agrees with BT's result of the global AdS, the instanton action accounts for the decay rate in the global AdS in which there are two black holes.\footnote{This is despite the fact that the dS$_3$ slice covers only part of the global AdS.}

The discussion beyond the probe limit goes as much in the same way as in the dS case. The instanton action \eqref{instantonactionbackreaction} for the AdS space can be obtained by the analytic continuation $H^{-1}\to iR_{AdS}$: 
\begin{align}
\Delta S_{E}=-\half(2\Delta S_{AdS})-{4\pi^2\over \kappa}\left[R_{AdS,{\rm i}}^{2}\sqrt{1+(R/R_{AdS, {\rm i}})^2}
-R_{AdS, {\rm o}}^{2}\sqrt{1+(R/R_{AdS, {\rm o}})^2}\right]\ ,
\end{align}
where $R=R_0({\rm AdS})$ in \eqref{radiusAdS}. As in the dS case, the second difference term can be identified with the difference of the thermal energy contributions $\beta_{\rm i} (2{\cal E}_{{\rm i}})-\beta_{\rm o} (2{\cal E}_{\rm o})$ with
\be
{\cal E}_{\rm i, o}=\half({\cal E}_{AdS, th})_{\rm i,o}\sqrt{g_{00}(R)_{\rm i, o}}\ ,
\ee
where the finite part of the thermal energy ${\cal E}_{AdS, th}=-R_{AdS}/(2G)$ and the temporal component $g_{00}(R)$ is the one of the global AdS metric which can be obtained from \eqref{staticpatch} by the analytic continuation $H^{-1}\to iR_{AdS}$. We thus have
\begin{align}
\Delta S_{E}=\half\left[-2\Delta S_{AdS}+\left(\beta_{\rm i}({\cal E}_{AdS, th})_{\rm i}\sqrt{g_{00}(R)_{\rm i}}-\beta_{\rm o}({\cal E}_{AdS, th})_{\rm o}\sqrt{g_{00}(R)_{\rm o}}\right)\right]\ .
\end{align}
As explained above, note again that both the entropy and the thermal energy are twice as much as those in the dS case.

\subsection{The case of de Sitter to Anti de Sitter}
\label{dStoAdSentropy}

As the third case we discuss when a de Sitter space decays and jumps into an Anti de Sitter space. Namely, it is when the outside cosmological constant $\Lambda_{\rm o}>0$ and the one inside is
\be
\Lambda_{\rm i}=\Lambda_{\rm o}-\kappa q\langle E\rangle < 0\ ,
\ee
where $\langle E\rangle =E_{\rm o}-{q\over 2}$ as defined before, and $R_{AdS,{\rm i}}^{-2}=-{\Lambda_{\rm i}\over 3}$ and $H_{\rm o}^2={\Lambda_{\rm o}\over 3}$. The change of the entropy is calculated as
\begin{align}
\Delta S=2S_{AdS, {\rm i}}-S_{dS, {\rm o}}={8\pi^2 q\langle E\rangle\over 3H_{\rm o}^4}+O(\kappa, q^2)\ .
\end{align}
Note that as remarked above, the factor of $2$ for the AdS entropy is to account for the fact that there are two topological black holes in the global AdS space. 

Meanwhile, since the outside spacetime is de Sitter, the instanton action relevant to this case is the one for the de Sitter space. In the probe limit this yields
\begin{align}
S_{M2}(\chi_0)=\frac{2 \pi ^2T_2}{H^3}-\frac{4\pi^2 qE }{3 H^4}+O(q^2)
=\half\left(\beta \Delta{\cal E}-\Delta S\right)+O(q^2)\ .
\label{ThermalInterpretationInstAction2}
\end{align}
Thus the decay rate correctly encodes the first law of thermodynamics for de Sitter and Anti de Sitter spaces.

Finally, the thermodynamic interpretation beyond the probe limit is a straightforward combination of the previous two cases. The instanton action in this case can be obtained from \eqref{instantonactionbackreaction} by the analytic continuation $H_{\rm i}^{-1}\to iR_{AdS, {\rm i}}$. We thus have
\begin{align}
\Delta S_{E}=\half\left[-(2S_{AdS,{\rm i}}-S_{dS, {\rm o}})+\left(\beta_{\rm i}({\cal E}_{AdS, th})_{\rm i}\sqrt{g_{00}(R)_{\rm i}}-{\beta_{\rm o}\over 2}({\cal E}_{th})_{\rm o}\sqrt{g_{00}(R)_{\rm o}}\right)\right]\ .
\end{align}

\section{Discussions and conclusions}
\label{Discussions}

We studied the instantons (or bounces) in the Brown-Teitelboim (BT) mechanism of relaxation of cosmological constant which is a cosmological version of the Schwinger mechanism.
In particular, we examined instanton solutions of the membrane nucleation (1) in the conformally flat metric of dS$_4$ space and (2) the dS$_3$ slice of 4d dS, flat and AdS spaces. The former manifests itself as being the most intuitive as false vacuum decay, while the latter is simpler and more suitable to discuss the BT mechanism. 
We first analyzed membrane instantons in the probe limit and then, as done in original BT's works, treat them as more fully-fledged gravitating domain walls to include the backreaction. We demonstrated that the latter reduces to the probe result as gravitational interactions are switched off, i.e. in the $\kappa\to 0$ limit.  As anticipated, the two analyses lead to essentially the same qualitative picture.

It is rather clear that the BT mechanism has some bearings  on (A)dS/CFT holography. To make this point sharper, as is the case with pair productions near black hole and de Sitter horizons, we showed that the BT instanton action for a membrane nucleation encodes the first law of thermodynamics of (A)dS space. In particular, the membrane instanton precisely accounts for the change of entropy of (A)dS space before and after nucleation, in good accordance with AdS/CFT in which the co-dimension one branes make up all degrees of freedom of AdS space. 

In light of the perspective our discussions and findings in the preceding sections might offer, we would like to end this section with (1) further remarks on (A)dS/CFT and (2) discussions on complexity.
 
\subsection{Comments on (A)dS/CFT}
\label{dS/CFT}

\begin{figure}[h]
\centering
\hspace{-.7cm}
\includegraphics[width=2.0in]{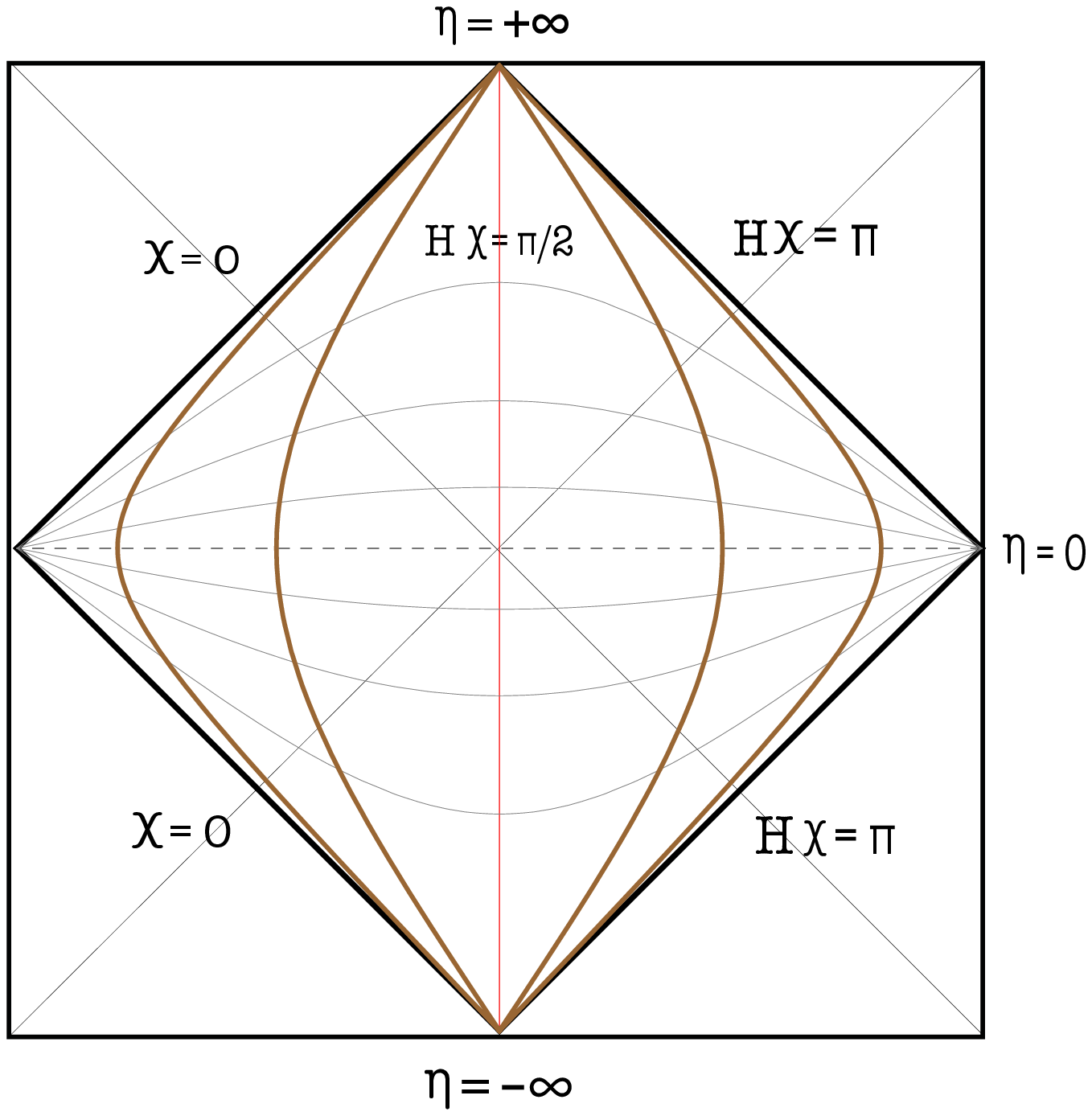}
\hspace{3.cm}
\includegraphics[width=1.1in]{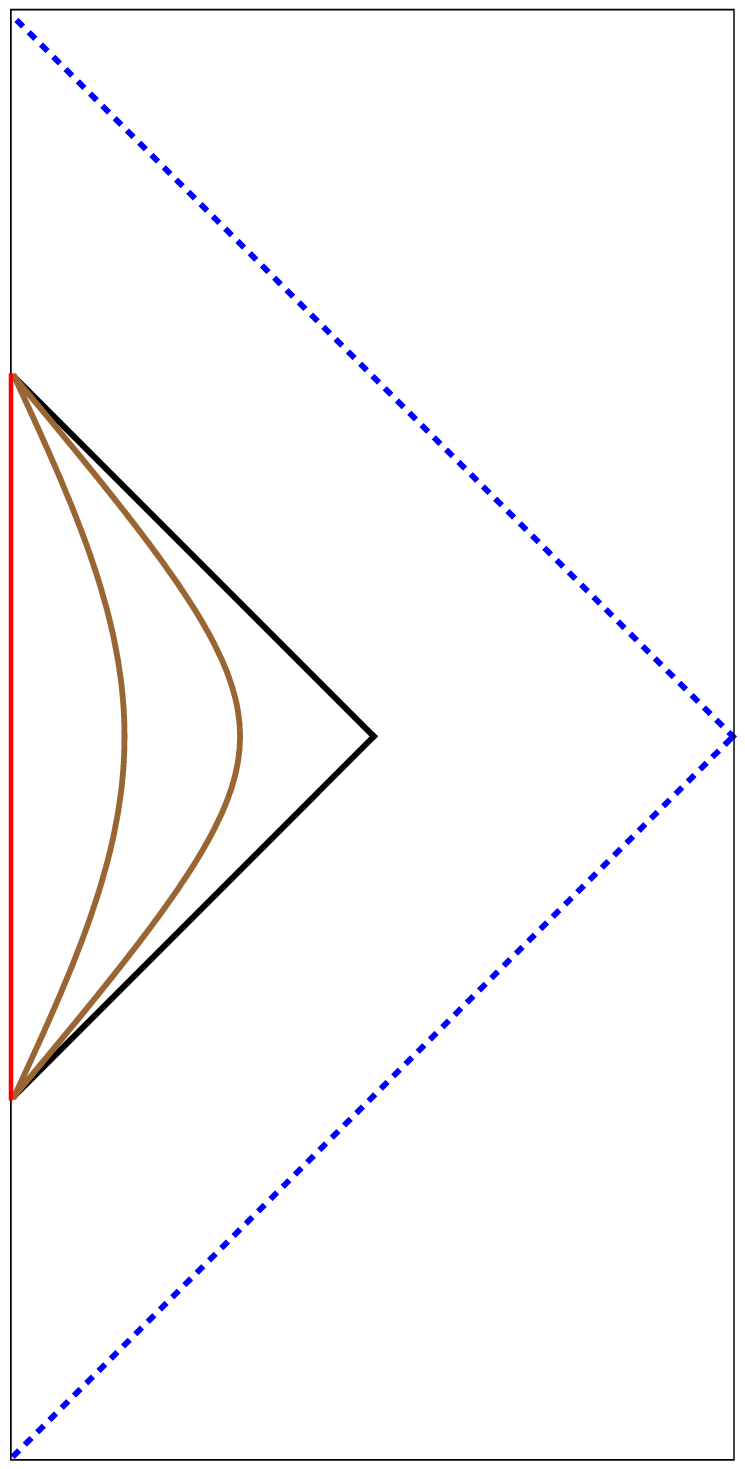}
\caption{The Penrose diagrams of de Sitter (left) and Anti de Sitter (right) spaces: (L) The dS$_3$ slice of dS$_4$, $ds_{\rm dS}^2=d\chi^2+H^{-2}\sin^2(H\chi)ds_{dS_3}^2$ in \eqref{dS3slicedS}, covers the diamond region bounded by black solid lines. The horizontal and vertical curves are constant $\eta$ and $\chi$ surfaces, respectively. The $H\chi=0, \pi$ and $\eta=\pm\infty$ constitute the cosmological horizons, i.e. the edges of the diamond. (R) The dS$_3$ slice of AdS$_4$, $ds_{\rm AdS}^2=d\rho^2+R_{AdS}^2\sinh^2(\rho/R_{AdS})ds_{dS_3}^2$ in \eqref{dS3slice}, covers the smaller triangle bounded by black solid lines inside the larger triangle of the Poincar\'e patch bounded by blue dotted lines. The vertical curves are constant $\rho$ surfaces.}
\label{fig:Penrose}
\end{figure}

As a membrane is nucleated, it feeds its entropy into the (A)dS space by just the right amount to create the new (A)dS space. This implies that membranes encode as many degrees of freedom as those of the (A)dS spaces. This nicely fits the idea of holography \cite{tHooft:1993dmi, Susskind:1994vu, Maldacena:1997re} and we wish to understand the relation between membrane nucleation and holography better. Let us first consider the AdS case. As remarked in the end of section \ref{dS3}, the critical flux corresponds to BPS branes, which was implied in \cite{Maldacena:1998uz}. Since the nucleation radius is infinity at the critical flux, the BPS branes are nucleated at the boundary of the AdS space (the red vertical line at the edge of the triangle in Figure \ref{fig:Penrose}) and there is only the inside spacetime bounded by the BPS branes, although the nucleation rate goes down to zero. Since there is only inside, all degrees of freedom can be accounted for by the BPS branes at the boundary. This picture has a strong resemblance to holography in which a holographic screen is located at the boundary. 

Let us now consider the dS case. There is no critical flux for de Sitter space. However, there is a maximal size at which membranes can be nucleated and it is the Hubble radius $R_H=H^{-1}$ and this corresponds to the limit of neutral charge $q=0$. Note that the decay rate does not become zero at this ``critical point'' unlike in the AdS case. Despite differences, being a maximal membrane, this is what might be the analogue of the BPS brane in the AdS case. In Figure \ref{fig:Penrose} the membrane at the Hubble radius corresponds to the red central vertical line in the diamond. In light of the above entropy argument, it is tempting to identify the membrane of the Hubble radius with a holographic screen. However, this timelike slice, though somewhat special, is not an asymptopia of de Sitter space and  it is hard to associate this slice with the natural domain to define meta-observables in de Sitter space. 
Moreover, if the CFT lives there, the symmetry would be $SO(2,3)$ instead of $SO(1,4)$. These are the reasons why the CFT is postulated to live on the spacelike surface at the future or past infinity in the dS/CFT conjecture \cite{Strominger:2001pn, Witten:2001kn}. 
Regardless of the understanding of the microscopic origin, the dS/CFT correspondence, as defined by an analytic continuation from AdS/CFT,  has proven to be a computationally useful and insightful approach to cosmology \cite{Maldacena:2002vr, McFadden:2009fg}.

We now wish to have a closer look at this analytic continuation in connection to the membrane nucleation discussed in Section \ref{dS3}.
We consider the dS$_3$ slice of AdS$_4$. The analytic continuation is performed as $R_{AdS}=iH^{-1}$ and $\rho=it +\beta/4$ ($\beta=2\pi/H$) together with the Euclidean continuation $\eta=i(\theta-\pi/2)$ of the dS$_3$ to the $S^3$:
\be
ds_{\rm AdS}^2=d\rho^2+R_{AdS}^2\sinh^2(\rho/R_{AdS})ds_{dS_3}^2\quad\longrightarrow\quad
ds_{\rm dS}^2=-dt^2+H^{-2}\cosh^2(Ht)ds_{S^3}^2
\ee
which is the closed de Sitter universe, i.e. the global dS.
In this Lorentzian spacetime, we look for a Euclidean $S^3$ membrane. The analysis is very similar to that in Section \ref{dS3} and the probe action yields
\begin{align}
\hspace{-.7cm}
S_{M2} = \int {dV_{S^3}\over H^3}\left[T_2H\cosh^2(Ht)\sqrt{{\cosh^2(Ht)\over H^2}-t'^2}+{qE\over 3H}\left(\sinh^3(Ht)+3\sinh(Ht)+C_0\right)\right]
\end{align}
where $t'=dt/d\theta$ and the constant $C_0$ is the gauge ambiguity of the 3-form potential.
The effective potential
\be
V_{\rm eff}(t)={1\over H^3}\left[T_2\cosh^3(Ht)+{qE\over 3H}\left(\sinh^3(Ht)+3\sinh(Ht)+C_0\right)\right]
\ee
has a minimum at 
\be
\coth(Ht_0)=-{3T_2H\over qE}\quad\Longrightarrow\quad
R=H^{-1}\cosh(Ht_0)=\frac{{3T_2\over |q|E}}{\sqrt{\left({3T_2H\over |q|E}\right)^2-1}}\ .
\ee
We note that this is not a bubble nucleation for false vacuum decay but rather a different kind of instanton. In fact, this is literally an instanton in the sense that it exists only at an instant of time $t=t_0$ and can be thought of as an example of S-branes which might have an interpretation as (real-time) instantons on unstable non-BPS branes \cite{Gutperle:2002ai}. Since it is a source of flux, the 4-form flux and the cosmological constant jumps across this membrane instanton.

The instanton action can be evaluated to
\be
S_{M2}(t_0)={2\pi^2|q|E\over 3 H^4}\frac{c^2-2+{\rm sign}(q)C_0\sqrt{c^2-1}}{\sqrt{c^2-1}}\ ,
\ee
where $c={3T_2H\over |q|E}$. The small $q$ expansions yield
\begin{align}
S_{M2}(t_0)={2\pi^2T_2\over H^3}+{2\pi^2C_0 {\rm sign}(q)qE\over 3H^4}+O(q^2)\ .
\end{align}
The choice of gauge, $C_0=-2\,{\rm sign}(q)$, yields the same result as \eqref{ThermalInterpretationInstAction} to this order. In fact, this would be the choice when analytically continued from AdS. 
With this gauge choice, the entropy story goes similar to the BT instantons and now there is a critical flux for the dS case at $c=1$, i.e.
\be
{T_2\over |q|}={E\over 3H}\ .
\ee 
The membrane radius $R$ is infinity and the membrane is located at the spacelike boundary at the future or past infinity. Now the parallelism is stronger and this is indeed very much analogous to the BPS branes in the AdS case.


\subsection{Complexity?}
\label{sec:complexity}

We now discuss that the {\it Lorentzian} bounce action might be closely related to computational complexity advocated by Susskind \cite{Susskind:2014rva}. In our de Sitter case, although there is no black hole, there are cosmological horizons and thus one may anticipate physics of complexity similar to that of black holes. 
The Lorentzian action \eqref{M2dS3} for a finite time interval $\eta$ yields
\begin{align}
S_{M2,L}(\chi_0,\eta)=-{2\pi\eta+\pi\sinh(2\eta)\over 2\pi^2}S_{M2}(\chi_0)\ ,
\end{align}
where $S_{M2}(\chi_0)$ is the Euclidean action \eqref{InstactiondS3} and  the only difference between the Lorentzian and Euclidean actions comes from the volume factors $V_{dS_3}(\eta)=2\pi\eta+\pi\sinh(2\eta)$ and $V_{S^3}=2\pi^2$. 
Note that the Lorentzian action increases as it approaches the cosmological horizon at $\eta=+\infty$.
In terms of the time
\be
H t=4\eta+2\sinh(2\eta)\ ,
\ee
the Lorentzian action takes the form
\begin{align}
S_{M2,L}(\chi_0,t)={t\over 2\pi} {-2S_{M2}(\chi_0)\over R_H}={t\over 2\pi} {\Delta S_{dS}-\beta\Delta{\cal E}\over R_H}\ ,
\end{align}
where we used \eqref{ThermalInterpretationInstAction}. It is then tempting to identify the Lorentzian bounce action with a change of complexity (up to the thermal energy),\footnote{The energy part is simply the dynamical time evolution since the amplitude $e^{iS_{M2,L}(\chi_0,t)}=e^{i\Delta{\cal C}}e^{- i \Delta{\cal E} t}$.}
\begin{align}
S_{M2,L}(\chi_0,t)+\Delta{\cal E} t \equiv\Delta{\cal C}= {t\over 2\pi} {\Delta S_{dS}\over R_H}\ ,\label{complexity}
\end{align}
since the complexity as defined in \cite{Susskind:2014rva} is of the form ${\cal C}={t\over 2\pi}{S_{\rm{\scriptscriptstyle BH}}\over R_{{\rm{\scriptscriptstyle BH}}}}$.
At the scrambling time $t_{\star}=2\pi R_H\ln S_{dS}$ \cite{Sekino:2008he}, this becomes
\be
\Delta{\cal C}_{\star}\simeq \Delta S_{dS}\ln S_{dS}
\ee
which is indeed a small change of the complexity at the Planckian layer, ${\cal C}_{\star}=S_{dS}\ln S_{dS}$, for a large $S_{dS}$.

Qualitatively, this discussion in the probe limit all carries over to the Lorentzian bounce action for the more fully-fledged BT's gravitating domain wall. The {\it Lorentzian} continuation of the BT instanton action \eqref{BTinstantonaction} is the on-shell gravity action (with the background subtraction) evaluated in a part, i.e. some finite time interval, of the Wheeler-DeWitt patch for $t=0$, as can be seen from Figure \ref{fig:Penrose}. Thus there is some resemblance to the conjecture made in \cite{Brown:2015bva}. 

In the Anti de Sitter case, the above argument remains much the same with $S_{dS}$ in \eqref{complexity} being replaced by $2S_{AdS}$, twice the entropy of the hyperbolic black hole \eqref{topBH}. However, the Lorentzian continuation of the BT instanton action \eqref{BTinstantonaction} with $H^{-1}=iR_{AdS}$ evaluated on the dS$_3$ slice does not seem to have any connection to the Wheeler-DeWitt patch, as can be seen from Figure \ref{fig:Penrose}. Nevertheless, we note that the dS$_3$ slice, to be more precise, two copies of them, do contain spacelike surfaces which pass through the interior of the hyperbolic black hole, since the coordinates of \eqref{dS3slice} and \eqref{topBH} have the relation
\be
r/R_{AdS}=\sqrt{1+\sinh^2(\rho/R_{AdS})\left(1-(\cosh\eta\sin\theta)^2\right)}\ ,
\ee 
where $\theta$ is the latitude angle of the $S^2$ membrane in the global dS$_3$. To elaborate, for a given angle $\theta$ the later time region defined by $\cosh\eta\sin\theta>1$ goes behind the horizon of the hyperbolic black hole, $r<R_{AdS}$. Except for the north and south poles $\theta=0, \pi$, the dS$_3$ slice always makes excursions into the interior. At the equator $\theta=\pi/2$ of the membrane, all surfaces except for $\eta= 0$ are inside the hyperbolic black hole. Note that the latitude $\theta$ is integrated (from $0$ to $\pi$) in the bounce action.
Thus it seems possible in principle that the Lorentzian bounce action measures some degree of complexity. 

In the real black hole case, it is conceivable that the Lorentzian action for a pair creation near the horizon, along the line of \cite{Parikh:1999mf}, may measure some degree of complexity and might have an interpretation as a small variation of complexity. It would be interesting to investigate further on this issue.


\section*{Acknowledgment}

We would like to thank Robert de Mello Koch, Vishnu Jejjala, Tetsuya Shiromizu and Joan Simon for discussions and comments. 
The work of SH was supported in part by the National Research Foundation of South Africa and DST-NRF Centre of Excellence in Mathematical and Statistical Sciences (CoE-MaSS).
Opinions expressed and conclusions arrived at are those of the author and are not necessarily to be attributed to the NRF or the CoE-MaSS.


\end{document}